\documentclass[%
 reprint,
 superscriptaddress,
 amsmath,amssymb,
 aps,
 prb,
 floatfix,
]{revtex4-2}

\usepackage[htt]{hyphenat}
\usepackage{graphicx}
\usepackage{dcolumn}
\usepackage{bm}
\usepackage[utf8]{inputenc}
\usepackage[T1]{fontenc}
\usepackage{mathptmx}
\usepackage{textcomp}
\usepackage{siunitx}
\usepackage{bigints}
\usepackage{upgreek}
\usepackage{hyperref}
\usepackage[switch]{lineno}
\usepackage[table,xcdraw,dvipsnames]{xcolor}
\usepackage{subcaption}
\usepackage{booktabs}
\usepackage{float}
\usepackage{color}
\usepackage[ruled]{algorithm2e}
\usepackage{multirow}
\usepackage{placeins}
\usepackage{url}
\usepackage{dsfont}
\usepackage{listings}

\definecolor{vsmod}{HTML}{136352}
\definecolor{vskey}{HTML}{707012}
\definecolor{vsvar}{HTML}{315b70}
\definecolor{vsstr}{HTML}{694d40}
\definecolor{vsimp}{HTML}{75104b}
\definecolor{vsmis}{HTML}{0f2845}
\definecolor{vscom}{HTML}{2d6614}
\definecolor{vsbac}{HTML}{ffffff}

\lstdefinelanguage{PythonEX}[]{}{
  morecomment  = [l]{\#},
  morestring   = [b]",
  morekeywords = [1]{,key,uniform,gen_hierarchy,eval_potential,at,get,shape,balanced_tree,arange,argmax,nanmax,nanmin,argpartition,pad,reshape,sort,jit,}, 
  morekeywords = [2]{,@partial,jax,range,jnp,lax,jaxfmm,random,}, 
  morekeywords = [3]{,def,None,}, 
  morekeywords = [4]{,from,import,as,for,in,return,}, 
}

\lstdefinestyle{myPython}{
    language     = PythonEX,
    basicstyle   = \bfseries\footnotesize\ttfamily\color{vsvar},
    frame        = single, 
    rulecolor    = \color{gray},
    commentstyle = \color{vscom},
    keywordstyle = \color{vskey},
    keywordstyle = [2]\color{vsmod},
    keywordstyle = [3]\color{vsmis},
    keywordstyle = [4]\color{vsimp},
    stringstyle  = \color{vsstr},
    emphstyle    = \color{pink}\underbar,
}

\newcommand*\diff{\mathop{}\!\mathrm{d}}
\renewcommand{\vec}[1]{\ensuremath{\boldsymbol{#1}}}
\newcommand{\norm}[1]{\left\lVert#1\right\rVert}
\DeclareMathOperator{\sgn}{sgn}

\begin{document}

\preprint{APS/123-QED}

\title{jaxFMM: An Adaptive, GPU-Parallel Implementation of the Fast Multipole Method in JAX}

\author{Robert~Kraft}    
\affiliation{Physics of Functional Materials, University of Vienna, Vienna, Austria}
\affiliation{Vienna Doctoral School in Physics, University of Vienna, Vienna, Austria}
\author{Florian~Bruckner}    
\affiliation{Physics of Functional Materials, University of Vienna, Vienna, Austria}
\author{Dieter~Suess}    
\affiliation{Physics of Functional Materials, University of Vienna, Vienna, Austria}
\author{Claas~Abert}    
\affiliation{Physics of Functional Materials, University of Vienna, Vienna, Austria}
\date{\today}

\begin{abstract}
We introduce jaxFMM, an open-source, adaptive, highly parallel point-charge Fast Multipole Method implementation for the Laplace kernel written in JAX. It is based on a non-uniform refinement strategy with on-the-fly rotation-based transforms tailored around JAX's just-in-time compiler, which results in extremely concise and simple code. Benchmarks show that the algorithm performs well at moderate accuracies, even for highly non-uniform charge distributions. JaxFMM already massively speeds up stray-field computations in micromagnetics and with JAX features like autodiff, novel applications such as inverse-design problems and machine-learning tasks can be tackled with ease in the future.

\end{abstract}

\maketitle


\section{Introduction}
\label{sec:introduction}

Since its inception in 1987 \cite{greengardFastAlgorithmParticle1987}, the Fast Multipole Method (FMM) has been successfully used to massively speed up a variety of numerical tasks. This ranges from the particle simulations the algorithm was initially designed to solve, to convolutions with arbitrary kernels via kernel-independent volume FMM formulations \cite{yingKernelindependentAdaptiveFast2004} or even purely algebraic use cases such as fast matrix multiplication \cite{yokotaFastMultipoleMethod2015}. Implementations of the algorithm must maintain a fine balance between far- and near-field interactions to maximize performance for a given accuracy. Furthermore, while the FMM can be parallelized efficiently, adaptive variants in particular require careful consideration to remain efficient. As a result, performant yet flexible FMM codes are often rather involved and difficult to develop. Notable examples here include pvfmm \cite{malhotraPVFMMParallelKernel2015}, exafmm \cite{wangExaFMMHighperformanceFast2021} and scalfmm \cite{blanchardScalFMMGenericParallel2015}.

With the rise of machine learning, Python libraries such as pytorch \cite{paszkePyTorchImperativeStyle2019} or JAX \cite{bradburyJAXComposableTransformations2018} seek to offer efficient GPU-parallel execution with comparatively little coding effort. Increasingly, this is also exploited for non-machine-learning tasks (see e.g. the "Awesome JAX" \cite{AwesomeJAXGithub2020} list of software using JAX). Another benefit of such frameworks is the availability of automatic differentiation, which can be used to compute gradients for inverse-design problems. In this work, we detail the implementation of jaxFMM, an adaptive point charge FMM library for the Laplace kernel written in JAX. Through JAX's just-in-time compilation, only roughly 800 lines of code yield a flexible, adaptive and GPU-parallel implementation of the FMM.

We begin with a brief introduction of the FMM and proceed by describing the unique aspects of our algorithm at each FMM stage. Then, we discuss JAX-specific implementation choices and their implications before we finish with benchmarks and a simple inverse-design example.

\section{Algorithm and Implementation}
\label{sec:methods}

\begin{figure}[!htbp]
    \centering
    \includegraphics[width=\columnwidth]{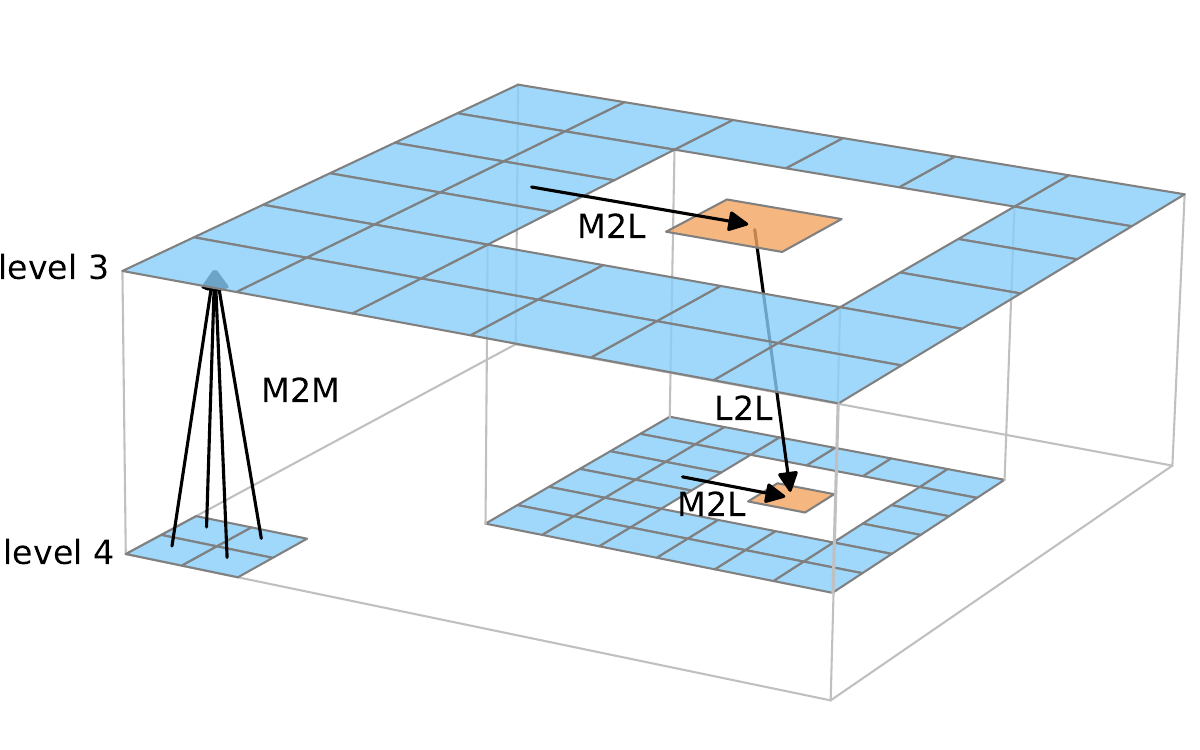}
    \caption{Schematic 2D-FMM setup: Combining multipoles (blue) into parent via M2M, adding them to locals (orange) via M2L and distributing locals to children via L2L.}
    \label{fig:fmm_schematic}
\end{figure}

The Fast Multipole Method approximates a potential $u$ arising from the convolutional integral over a domain $\Omega$,
\begin{align}
    \label{eq:convolution}
    u(\vec{r}) = \int_\Omega K(\vec{r} - \vec{r}') \rho(\vec{r}') \diff \vec{r}',
\end{align}
where $K$ is referred to as the kernel and $\rho$ is the source density. Here, the underlying idea of the algorithm is to hierarchically form and recombine expansions which approximate the potential caused by sources at long distances. More specifically, so-called multipole expansions are formed and combined into local expansions with multipole-to-multipole (M2M), multipole-to-local (M2L) and local-to-local (L2L) transformation operators wherever the prescribed accuracy permits (Fig. \ref{fig:fmm_schematic}). Then, the local expansions are used to compute far-field contributions to the potential, which drastically decreases the total amount of interactions that must be computed. JaxFMM currently only supports solving Poisson's equation for point charges, i.e.
\begin{align}
    \label{eq:laplace_kernel}
    K (\vec{r}, \vec{r}') &= -\frac{1}{4\pi} \frac{1}{\lVert \vec{r} - \vec{r}' \rVert}, \\
    \label{eq:poisson_density}
    \rho (\vec{r}') &= \sum_{i=1}^N q_i \cdot \delta (\vec{r}'-\vec{r}_i),
\end{align}
where $N$ charges $q_i$ are placed at positions $\vec{r}_i$ and $\delta$ is the Dirac delta function.

\subsection{Hierarchy Generation}
\label{subsec:hierarchy}

The original formulation of the FMM uses a uniform refinement strategy for its hierarchy, i.e. particles are assigned to boxes which are always cubes and split into 8 equally sized children. As a result, the algorithm is not adaptive: For the same number of charges $N$, non-homogeneous charge distributions can increase the computational effort significantly since the number of charges contained in each leaf box at maximum tree depth will vary considerably. 

\begin{figure}[!htbp]
    \centering
    \includegraphics[width=\columnwidth]{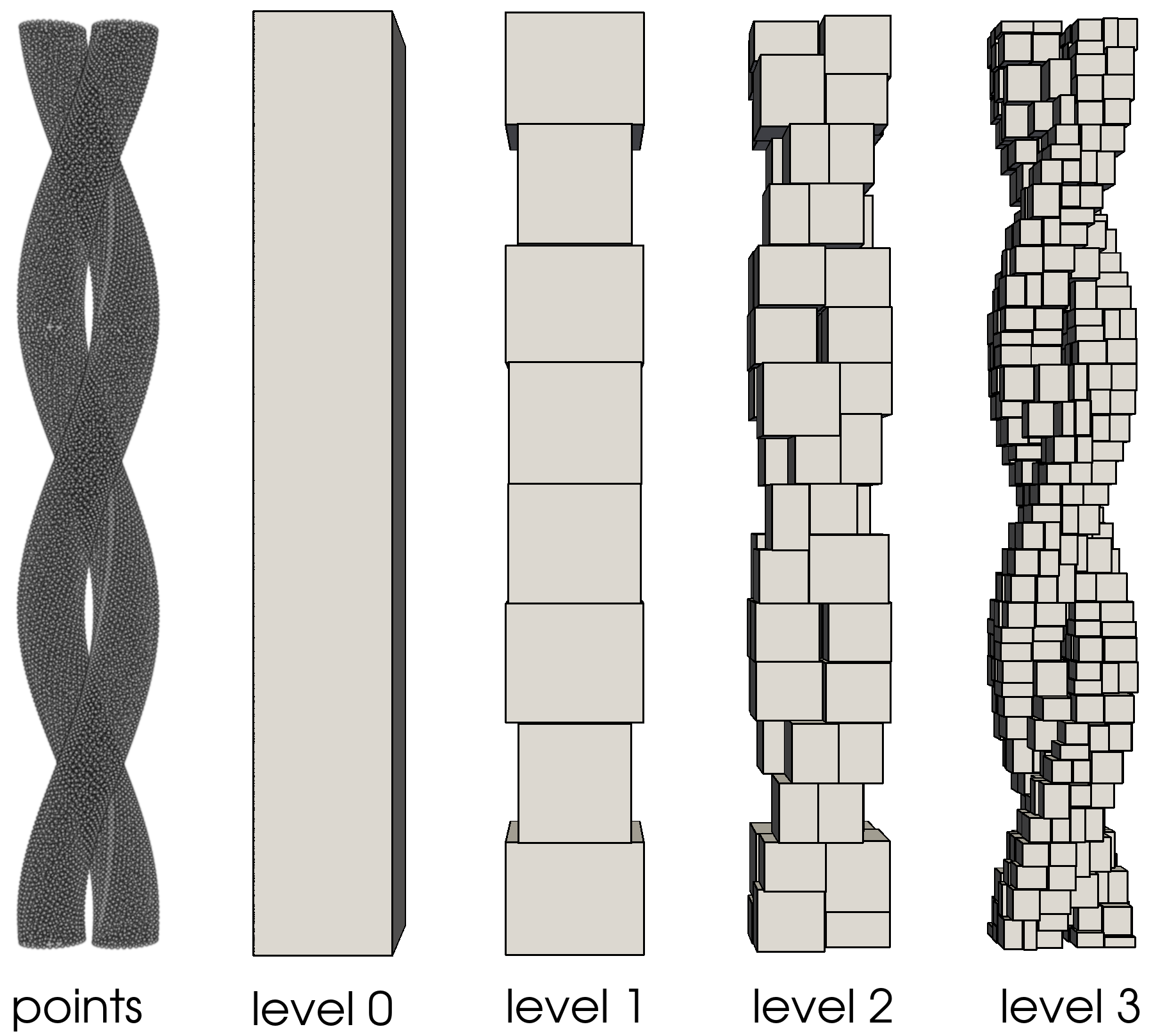}
    \caption{From left to right: Double helix point distribution, jaxFMM hierarchy levels $0-3$ ($s=3$ splits per box).}
    \label{fig:DNA_hierarch}
\end{figure}

To counter this increase in cost for such geometries or non-uniform charge distributions, we follow a non-uniform splitting approach first published in \cite{engblomWellseparatedSetsFast2010}: Instead of requiring strictly cubic simulation boxes, we allow rectangles of arbitrary shape and size and split not in the geometric center, but at the median value of the coordinates. More specifically, boxes are subdivided as follows:
\begin{enumerate}
    \item Find the longest side of the box along the principal axes by examining maxima and minima of the particle positions in the box.
    \item Split the box perpendicular to the corresponding axis direction by finding the median value of the particle coordinates on this axis and partitioning the position array into a half lower than the median and a half greater than the median.
    \item Repeat the above procedure for each half. The depth of this recursion is limited to a user-defined number of splits $s$ per level, such that each box has exactly $2^{s}$ children.
    \item Generate levels until each box holds no more than $N_\text{max}$ particles.
\end{enumerate}
As can be seen above, we permit an arbitrary number of children per box and split perpendicular to the longest side of the box. The parameters $s$ and $N_\text{max}$ are used to balance the cost of near- and far-field interactions: Increasing $N_\text{max}$ or $s$ decreases tree depth and results in a higher number of near-field interactions. This increase in near-field cost is accompanied by a higher accuracy, since the near-field is computed analytically. Additionally, $s$ controls the lower bound of the number of particles per box $N_\text{box}$, $N_\text{max}/2^s < N_\text{box} \leq N_\text{max}$, which has similar implications on accuracy and performance. Note that $s$ could be set individually on each level, but is set globally in jaxFMM, i.e. every box is split into the same number of children. Figure \ref{fig:DNA_hierarch} shows an example hierarchy generated by jaxFMM.

The non-uniform splitting algorithm in this form has the following advantages/disadvantages:
\begin{itemize}
    \item[$+$] Straightforward implementation.
    \item[$+$] The resulting $2^{s}$-tree hierarchy is \emph{balanced}, i.e. boxes at the same depth contain almost exactly the same number of charges. In fact, boxes contain either $\lceil N/2^{l \cdot s} \rceil$ (ceiling) or $\lfloor N/2^{l \cdot s} \rfloor$ (floor) particles on level $l$.
    \item[$+$] As a byproduct of the hierarchy generation process, the data is intrinsically arranged like a Z-order curve, which improves data locality \cite{mortonComputerOrientedGeodetic1966}.
    \item[$\sim$] $\mathcal{O}(N \log N)$ average complexity: $\mathcal{O}(\log N)$ levels with one $\mathcal{O}(N)$ average complexity partition each.
    \item[$-$] Interaction lists must be generated and stored.
    \item[$-$] Boxes may have poor aspect ratio in some cases, increasing the number of interactions that must be computed.
    \item[$-$] Efficiently precomputing transformation operators is no longer possible since box positions are not on a regular grid.
\end{itemize}
Overall, compared to other adaptive FMM variants, the non-uniform splitting is very simple to implement whilst still yielding balanced trees. However, at the same time it loses access to the optimizations that stem from the regular grid in uniform refinement methods. For example, storing precomputed transformation operators has a very large memory overhead since operators must be stored for every single interaction compared to a relatively small set of operators in uniform refinement methods. When fast routines for computing the operators are unavailable, computing them on-the-fly may be prohibitively expensive.

\subsection{Interaction Lists}

For the generation of local expansions in the leaf boxes at maximum depth, the FMM uses the multipole expansions of lowest possible depth that still satisfy an error bound to save computational cost. In the case of a uniform refinement FMM, the interaction partners of each box are known ahead of time, since the shape of the boxes and their relative positions are always the same (see e.g. the repeating pattern in Fig. \ref{fig:fmm_schematic}). Non-uniform splitting however requires the computation of interaction lists. To determine if two boxes are well-separated, we employ the criterion from \cite{engblomWellseparatedSetsFast2010},
\begin{align}
    \label{eq:wellsep}
    R + \theta r \leq \theta d,
\end{align}
where $\theta$ is a parameter controlling the error set by the user, $R$ refers to the radius of the larger box, $r$ to the radius of the smaller box and $d$ to the distance between their centers. We use this condition to generate interaction lists as follows:
\begin{enumerate}
    \item Starting on the root level, compare every box with all other boxes using \eqref{eq:wellsep}. Every box that fulfills the criterion gets added to the list of weakly coupled boxes on this level. The interactions of weakly coupled boxes will be computed via expansions later on.
    \item Proceed to the next level and repeat the above, but ignore the children of all well-separated parent boxes, as indicated by the previously generated interaction list. Do this until reaching the leaf level.
    \item At the leaf level, some boxes that are not weakly coupled will remain. They are added to a list of strongly coupled boxes, whose interactions will be computed directly later on.
\end{enumerate}
In \cite{engblomWellseparatedSetsFast2010}, it is shown that the relative error $\varepsilon$ of the $p$-th order FMM solution $u_p$ compared to the analytic potential $u$ then is bounded as follows:
\begin{equation}
    \label{eq:errbnd}
    \varepsilon = \left|\frac{u - u_p}{u} \right| \leq C \frac{\theta^{p+1}}{(1-\theta)^2}\,,
\end{equation}
where $C$ is a constant.
By examining the geometry of a nearest-neighbor uniform FMM splitting, we see that $\theta=\sqrt{3}/(4-\sqrt{3})\approx 0.7637$ in this case, which is why we choose a similar value ($\theta=0.77$) as the default in jaxFMM.

\begin{figure}[!htbp]
    \centering
    same-level interactions
    \includegraphics[width=\columnwidth]{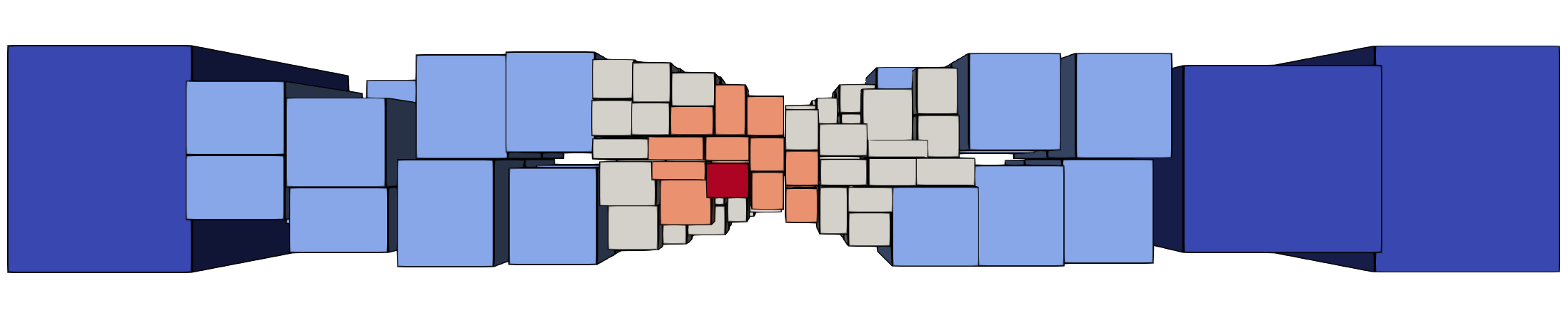}
    cross-level interactions
    \includegraphics[width=\columnwidth]{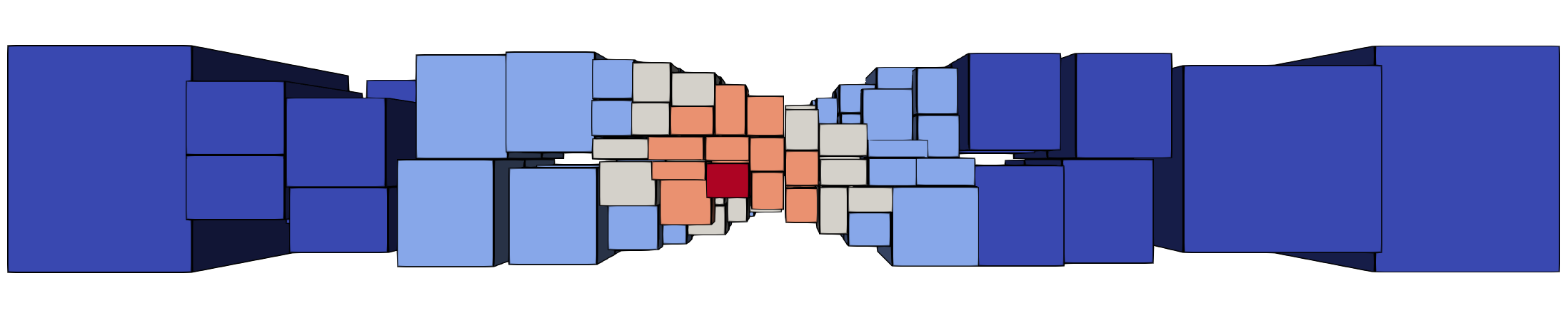}
    \caption{Same-level (top) vs cross-level (bottom) interaction lists, for a given box (red): Strong coupling (orange), weak coupling at decreasing target level $l_\text{trg}$ (white -- dark blue).}
    \label{fig:DNA_wellsep}

    \begin{tabular}{|cc|r|}
        \hline
        \multicolumn{1}{|c|}{$l_\text{src}$} & $l_\text{trg}$ & \multicolumn{1}{c|}{interactions} \\ \hline
        \multicolumn{1}{|c|}{1}      & 1      & 30                              \\ \hline
        \multicolumn{1}{|c|}{2}      & 2      & 1380                             \\ \hline
        \multicolumn{1}{|c|}{3}      & 3      & 40072                            \\ \hline
        \multicolumn{2}{|c|}{Total:}          & 41482                            \\ \hline
        \end{tabular}
        \quad vs \quad
        \begin{tabular}{|cc|r|}
        \hline
        \multicolumn{1}{|c|}{$l_\text{src}$} & $l_\text{trg}$& \multicolumn{1}{c|}{interactions} \\ \hline
        \multicolumn{1}{|c|}{1}      & 1      & 30                               \\ \hline
        \multicolumn{1}{|c|}{1}      & 2      & 96                              \\ \hline
        \multicolumn{1}{|c|}{2}      & 2      & 612                              \\ \hline
        \multicolumn{1}{|c|}{2}      & 3      & 2467                             \\ \hline
        \multicolumn{1}{|c|}{3}      & 3      & 20336                             \\ \hline
        \multicolumn{2}{|c|}{Total:}          & 23541                            \\ \hline
    \end{tabular}
    \label{tab:same_vs_cross}
    \captionof{table}{Total number of weakly coupled interactions for the double helix distribution. Same-level (left) vs cross-level (right) interactions.}
\end{figure}

The interaction list generation can be generalized further to allow separate sources and targets, i.e. evaluating the potential at positions different from the charge positions. This is useful whenever the potential at the source locations is not of interest, for example in the inverse design demonstration in \ref{subsec:inverse_demo} where the potential of the shielding charges needs only be known at a certain distance inside and outside the shielding region, but not in the shielding region itself. In this case, two hierarchies must be generated: one for the sources and one for the targets. Then, source boxes are compared with target boxes using \eqref{eq:wellsep} to generate the interaction lists. Note however that the number of levels in the source and target hierarchy are not necessarily the same and hence it no longer makes sense to only compare boxes on the same level. In other words, we now allow cross-level interactions such as weakly coupling boxes on source level $l_\text{src}=0$ to boxes on target level $l_\text{trg}=1$. While comparing all source levels with all target levels would certainly be optimal, it would also be prohibitively expensive. Therefore, we start at $l_\text{src}, l_\text{trg} = 0$ and then proceed as follows:
\begin{enumerate}
    \item Increase the level with larger average box radius.
    \item If the average radii are equal, increase the source level.
    \item If there are no more source levels, increase the target level and vice versa.
    \item Quit if all interactions have been accounted for or if there are no more levels available.
\end{enumerate}
The above algorithm aims to decrease $R$ in \eqref{eq:wellsep}, ensuring that levels with similar box sizes are compared. We can also apply this level-marching scheme in the case where the source and target points are the same, which decreases the total number of interactions that must be computed. For example, in Figure \ref{fig:DNA_wellsep}, we show all the source boxes considered for a given target box (red) in the double helix hierarchy of Figure \ref{fig:DNA_hierarch}. The boxes are colored by the target level at which they are included in the local expansion via M2L transformation, from level 1 (dark blue) to level 3 (off-white). Orange boxes are not included in the local expansion, their contribution to the potential is evaluated directly instead. Including boxes at the earliest possible target level avoids handling their children individually and thus decreases the number of interactions. Cross-level interactions facilitate this, which is clearly visible by the increase in dark and light blue boxes in Figure \ref{fig:DNA_wellsep} as well as the total interaction numbers for this example listed in Table \ref{tab:same_vs_cross}. We note that there is a decrease in data locality introduced by the cross-level interactions, but our level-marching scheme limits its effect since we increment levels one at a time and thus we observe performance improvements directly proportional to the reduction in interaction numbers. A similar, albeit less flexible performance improvement is also known from uniform refinement FMM \cite{gumerovFastMultipoleMethods2008}.

\subsection{Basis Functions}
\label{subsec:basis}

For the Laplacian kernel \eqref{eq:laplace_kernel}, expansions are commonly done in spherical coordinates $\vec{r} = (r,\theta,\varphi)$ using spherical harmonic functions $Y_n^m$:
\begin{align}
    \label{eq:laplace_expansion}
    K (\vec{r}, \vec{r}') &= \sum_{n=0}^\infty \sum_{m=-n}^n R^{-m}_n (\vec{r}') S^m_n (\vec{r}), \\
    \label{eq:regular_basis}
    \quad R^m_n(\vec{r}) &= r^{n} Y_n^m (\theta,\varphi), \\
    \label{eq:singular_basis}
    \quad S^m_n(\vec{r}) &= \frac{Y_n^m(\theta,\varphi)}{r^{n+1}}.
\end{align}
Here, we refer to $R^m_n$ as the regular basis and $S^m_n$ as the singular basis. Inserting \eqref{eq:poisson_density} and \eqref{eq:laplace_expansion} into \eqref{eq:convolution} and truncating the series at finite order $p$ leads to \emph{multipole expansions} of the form
\begin{align}
    \label{eq:mpl_series}
    u(\vec{r}) &\approx -\frac{1}{4\pi} \sum_{n=0}^{p} \sum_{m=-n}^n C^m_n S^m_n(\vec{r}),\\
    \label{eq:mpl_coeff}
    C_n^m &= \sum_{i=1}^N q_i R_n^{-m}(\vec{r}_i),
\end{align}
with multipole coefficients $C_n^m$.
It can be shown that the error of expansion \eqref{eq:mpl_series} scales as $(a/r)^{p+1}$ for evaluation positions $\vec{r}$ outside a ball with radius $a$ around the origin, where $a$ refers to the distance of the charge furthest from the origin \cite{greengardNewVersionFast1997}. Similarly, \emph{local expansions} are defined as
\begin{align}
    \label{eq:loc_series}
    u(\vec{r}) &\approx -\frac{1}{4\pi} \sum_{n=0}^{p} \sum_{m=-n}^n D^m_n R^m_n(\vec{r}), \\
    \label{eq:loc_coeff}
    D_n^m &= \sum_{i=1}^N q_i S_n^{-m}(\vec{r}_i),
\end{align}
with local coefficients $D_n^m$ and an error that scales like $(r/a)^{p+1}$ inside the ball with radius $a$, where $a$ now represents the distance of the charge closest to the origin.

The above expansions are based on complex-valued spherical harmonic basis functions (\ref{eq:regular_basis}, \ref{eq:singular_basis}). However, it is straightforward to show that $C^{-m}_n = \bar{C}^m_n$ and $D^{-m}_n = \bar{D}^m_n$, i.e. expansion coefficients of opposite sign in $m$ are complex conjugates. Therefore, it is possible to either only save half of the coefficients or to use a real-valued basis instead. As shown in \cite{gumerovFastMultipoleMethods2008}, recursion relations are available for cheaply computing a real-valued regular basis. We use slightly different recursions to not only compute our regular basis \eqref{sec:reg_rec}, but also the gradient of the regular basis \eqref{sec:reg_grad_rec}, which can be used to directly compute fields instead of potentials. We note that our basis differs slightly from the reference, such that multipole coefficients are computed and local expansions evaluated as follows:
\begin{align}
    C_n^m &= \sum_{i=1}^N q_i \hat{R}_n^{m}(\vec{r}_i), \\
    u_p(\vec{r}) &= -\frac{1}{4\pi} \sum_{n=0}^{p} \sum_{m=-n}^n (2-\delta_{0m}) \sgn(m) D^m_n \hat{R}^m_n(\vec{r}),
\end{align}
where $\delta_{0m}$ is the Kronecker delta and $\sgn$ the sign function including zero ($\sgn(0) = 1$). For completeness, we also include recursions to compute the corresponding singular basis \eqref{sec:sing_rec}.

\subsection{Transformations}
\label{subsec:transf}
As already mentioned, due to the non-uniform shape and position of the boxes in the hierarchy, precomputing transformation operators is too expensive in terms of memory. Operators must be computed on-the-fly for every single interaction instead of precomputing one, relatively small reoccurring set. JaxFMM uses $\mathcal{O}(p^3)$ rotation-based transformation operators, which rotate coefficients before transforming to avoid the $\mathcal{O}(p^4)$ scaling of the direct "nested-sum" operators. To compute the rotations quickly on-the-fly, we implemented fast Jacobi polynomial recursions from \cite{wangEffectiveEfficientAlgorithm2022a} (eqs. 13-14) in JAX for computing the necessary Wigner matrices. We note that due to symmetry, only roughly 1/4 of the Wigner matrix coefficients must be computed (\cite{wangEffectiveEfficientAlgorithm2022a}, eqs. 15a-15c). Then, we derived and use the real-basis equivalent of the rotations and diagonal transformations in \cite{dachselFastAccurateDetermination2006} (App. \ref{sec:transf_ops}).

\subsection{JAX Implementation}
\label{subsec:jax}

JAX is a Python library for large-scale array computations on accelerators (GPU/TPU), with a focus on machine-learning applications. It is similar to numpy in terms of usage, but its main appeal is the just-in-time (jit) compilation functionality: Functions with a jit-decorator are compiled on their first call, which introduces compilation overhead for said first call but results in much faster runtimes on subsequent calls. The optimizations by the compiler enable high performance for comparatively little coding effort. Additionally, for almost all jit-compiled functions, automatic differentiation by the input variables is available.

However, jit-compilation relies on static array sizes. In other words, if any of the input arrays changes in size, the function must be recompiled and functions returning outputs of dynamic size cannot be jit-compiled. Additionally, the execution logic of a jit-compiled function cannot depend on the values of its inputs. In the following, we describe how jaxFMM deals with these limitations.

First, we compute the FMM hierarchy. It is straightforward to write this process in a jit-compilable form, but we note that the splitting function must be recompiled whenever the number of input particles changes. Note that frequent recompilation can be avoided here by adding placeholder padding in the particle coordinate and charge arrays. Then, we pad the particle coordinates and charges sorted by leaf box, such that every leaf box contains exactly the same number of particles. Because the tree is balanced, only a very small amount of padding is inserted here (at most one particle per box).

Next, we determine the interaction lists. In the current version of jaxFMM, this process is not yet jit-compilable since it contains a while loop whose number of iterations depends on the input and it outputs interaction lists of dynamic size. However, setting a fixed buffer size for the output and a fixed tree depth can also make this step jit-compilable. Since this requires careful handling of the buffer size and tree depth, it has been avoided for now and will be added in a future update.

The returned interaction lists contain pairs of source and target indices across all levels of the tree for the M2L transformations and the direct near-field interactions on the leaf level, respectively. Therefore, since all levels are handled simultaneously, no further padding must be inserted and jit-compiling the M2L/direct evaluation stages is straightforward. Similarly, the perfect tree balance guarantees that M2M/L2L operations require no padding and can be jit-compiled with ease.

Overall, in the current jaxFMM version the setup cost for compiling the functions and building the tree must be paid every time the particle coordinates and/or number change. In return, we achieve static array sizes with minimal padding. For repeated potential evaluations on the same geometry, the hierarchy information can be cached and reused to evaluate the potential with different charges.

All computations take place on the same device, either the CPU or GPU but not both. Correspondingly, all data is either stored in RAM (CPU) or VRAM (GPU). To further optimize runtime and memory usage, we batch all routines such that the memory required for computing each batch roughly matches the L2-cache size of the processing unit.

As an example for jaxFMM code, Listing \ref{lst:unitcube} shows a script which computes the potential of charges uniformly distributed in a cube: After randomly generating the points and charges, a hierarchy information object \texttt{hier} is created which is then used to compute the potential of the generated charges. Furthermore, to illustrate the simplicity of both JAX and the non-uniform splitting approach, we also show the code that generates the $2^s$-tree hierarchy (Listing \ref{lst:hierarch}). Note that the fill value \mbox{\texttt{-jnp.nan**2}} is intentional, as \texttt{jax.argpartition} only sorts NaNs with negative sign to the front of the array.

\begin{lstlisting}[caption={Minimal uniform cube demo.},label=lst:unitcube]
from jax import random
from jaxfmm import *

### Generating points and charges
N = 2**15
rng = random.key(124)
pts = random.uniform(rng,(N,3))
chrgs = random.uniform(rng,N)

### Generating the FMM hierarchy
hier = gen_hierarchy(pts,p=2,theta=0.5)

### Computing the potential
pot_FMM = eval_potential(chrgs,**hier)
\end{lstlisting}

\begin{lstlisting}[caption={Code determining $2^s$ tree sorting. Inputs: Charge positions (\texttt{pts}), max. depth (\texttt{max\_l}) and splits per level (\texttt{s}). Output: Index array sorting points into leaves (\texttt{idcs}).},label=lst:hierarch]
import jax
import jax.numpy as jnp

@partial(jax.jit,static_argnames=["max_l","s"])
def balanced_tree(pts,max_l,s=3):
    n_chi = 2**s
    idcs = jnp.arange(pts.shape[0])[None,:]
    for l in range(max_l*s):
        splitpos = idcs.shape[1]//2
        needpad = idcs.shape[1]%2
        pts_srt = pts.at[idcs].get(mode="fill", fill_value=-jnp.nan**2)
        ax_split = jnp.argmax(jnp.nanmax(pts_srt,axis=1) - jnp.nanmin(pts_srt,axis=1),axis=1)
        idcs = idcs[jnp.arange(idcs.shape[0])[:,None],jnp.argpartition(pts_srt[jnp.arange(ax_split.shape[0]),:,ax_split],splitpos,axis=1)]
        idcs = jax.lax.pad(idcs,pts.shape[0],[(0,0,0),(0,needpad,0)])
        idcs = idcs.reshape((-1,idcs.shape[1]//2))
    idcs = jnp.sort(idcs,axis=1)
    return idcs
\end{lstlisting}

\section{Numerical Experiments}
\label{sec:results}

In the following section, we evaluate the performance of jaxFMM for a variety of charge distributions, hardware configurations and parameter settings. Computations take place on the Vienna Scientific Cluster VSC5, where GPU nodes are equipped with Nvidia A100 (40GB) cards and CPU nodes use AMD Epyc 7713 processors \cite{ASCUserDocumentation2022}. The GPU benchmarks are limited to one GPU, since distribution of data across multiple GPUs is not yet supported. Also, the CPU benchmarks use only one CPU core to investigate sequential performance. Finally, the jaxFMM version used for the benchmarks can be found under the \mbox{"\texttt{v0.2.0\_p}"} tag in the GitLab repository \cite{JaxFMMGitlabRepository2025}. 

\subsection{Uniform Cube and Sphere Shell}
\label{subsec:unitcube}

In this first experiment, we evaluate jaxFMM's performance for computing the potential of two different distributions: Points distributed uniformly in a cube and on the surface of a sphere. Particle numbers are varied from $N = 2^{11}$ to $N = 2^{27}$ in 16 even logarithmic steps, the number of splits per parent is set to $s=2$ and the three parameter sets in Tab. \ref{tab:parsets} are used. Said parameters showed the best performance in preliminary test runs for their respective accuracy.

\begin{table}[!htbp]
\begin{tabular}{|c|c|ccc|c|}
\hline
Set & Accuracy & $N_\text{max}$ & $p$ & $\theta$ & Precision \\ \hline
1)  & 3 digits & 128            & 5   & 0.70     & single    \\ 
2)  & 6 digits & 256            & 9   & 0.50     & single    \\ 
3)  & 9 digits & 256            & 10  & 0.27     & double    \\ \hline
\end{tabular}
\caption{Parameter sets used in the benchmarks, intended for 3,6 and 9 digits of accuracy, respectively.}
\label{tab:parsets}
\end{table}

Analytic potentials $u$ are also computed for $N\leq 2^{20}$ to determine the relative $L^\infty$-norm error $\epsilon$ of the FMM solution $u_\text{FMM}$,
\begin{equation}
    \epsilon = \frac{\norm{u - u_\text{FMM}}_\infty}{\norm{u}_\infty}.
\end{equation}
The reported runtimes are the minimum runtimes from 100 repeated potential evaluations. Finally, we note that some data points are missing for large $N$ due to running out of memory for temporary arrays. This issue is related to excess memory usage in the hierarchy assembly and will be addressed in a future version of jaxFMM.

\begin{figure}[!htbp]
    \centering
    \includegraphics[width=0.75\columnwidth]{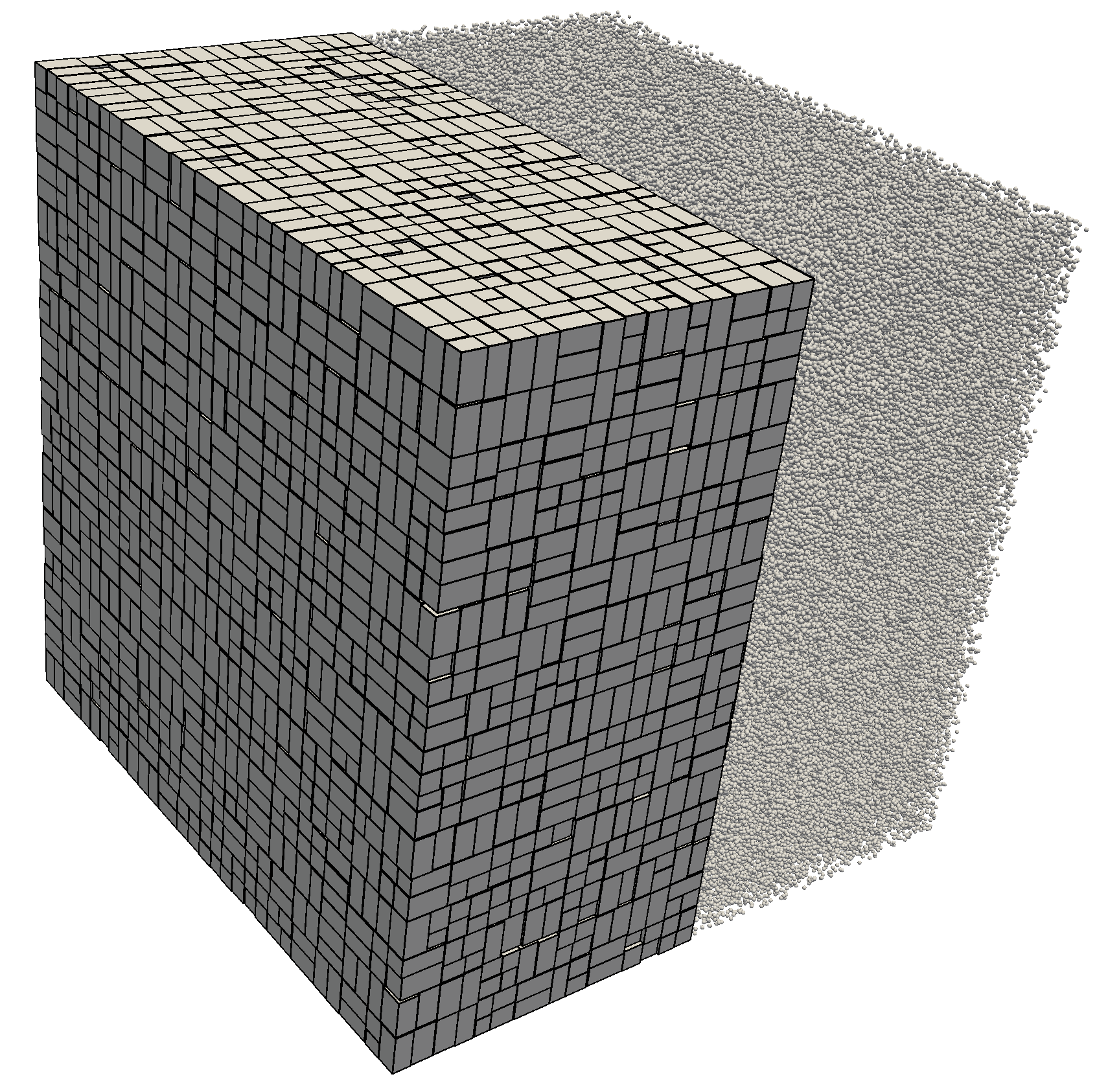}
    \caption{Level 7 of the uniform cube hierarchy (left) and source points (right) for $N=2^{20}$.}
    \label{fig:uc_hier}
\end{figure}

\begin{figure}[!htbp]
    \centering
    \includegraphics[width=\columnwidth]{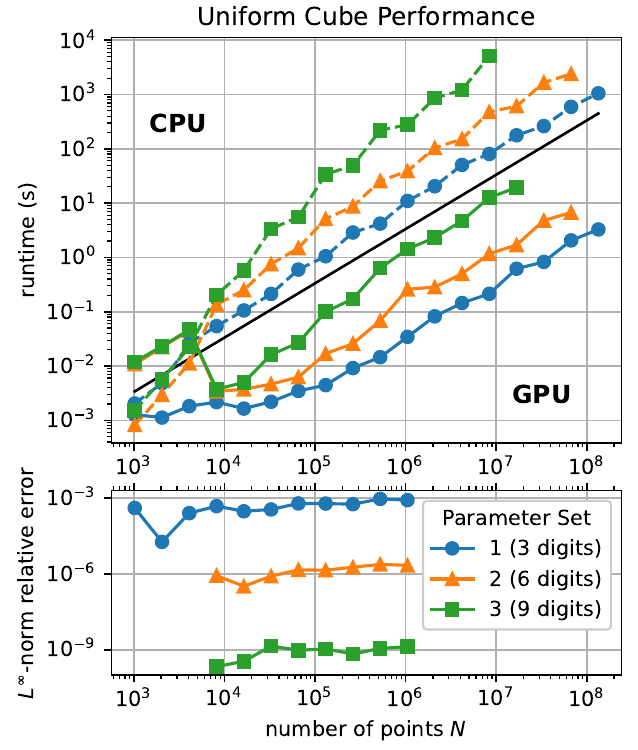}
    \caption{Uniform cube performance on CPU (dashed lines) and GPU (solid lines).}
    \label{fig:cube_perf}
\end{figure}

\begin{figure}[!htbp]
    \centering
    \includegraphics[width=\columnwidth]{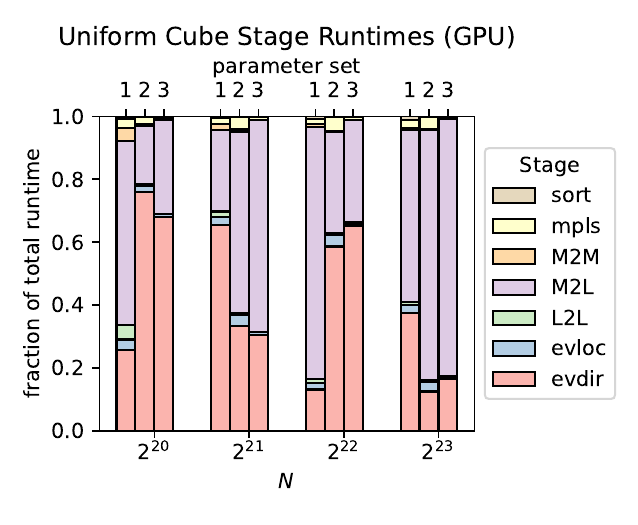}
    \caption{Uniform cube stage timings for select problem sizes.}
    \label{fig:cube_substep}
\end{figure}

First, we examine uniformly distributed charges in a cube, a good fit even for non-adaptive FMM variants. Figure \ref{fig:uc_hier} shows the generated distribution and hierarchy at level 7 for $N = 2^{20}$.
Runtimes and relative $L^\infty$-norm errors for different parameter settings on CPU and GPU are shown in Figure \ref{fig:cube_perf}. Overall, the implementation performs as expected: While the GPU resources are not yet exhausted, the dispatch overhead dominates and runtimes increase only slightly. Once the system is sufficiently large, runtimes scale linearly. Note the large gap in runtime between CPU and GPU execution, which indicates that JAX struggles to optimize the code for CPUs.
All parameter sets approximately reach their desired accuracy in the $L^\infty$-norm. We also show timings of the individual stages on the GPU (Figure \ref{fig:cube_substep}): Sorting and padding the points (sort), generating the initial multipole coefficients on the leaf level (mpls), ascending the hierarchy via M2M transforms (M2M), accounting for weakly coupled boxes via M2L transforms (M2L), descending the hierarchy via L2L transforms (L2L), evaluating the local expansions for the far field (evloc) and evaluating the remaining contributions directly for the near field (evdir). Higher accuracy computations come with increased computational cost, either via the $\mathcal{O}(p^3)$ cost of the far-field transformation operators, via an increased amount of M2L/direct interactions from decreasing $\theta$ or via an increased number of directly interacting points through a higher $N_{\text{max}}$. An interesting observation here is how the number of particles per box affects stage timings. For example, parameter sets 2 and 3 contain $N_\text{box}=(256,128)$ particles per box at $N=(2^{22}, 2^{23})$. The ratio of near-field (evdir) to far-field (all other stages) runtimes changes accordingly. Generally however, the cost for far- and near-field interactions is relatively balanced and was successfully fine-tuned not only by adjusting $p$, $\theta$ and $N_\text{max}$ but also the number of splits $s$.

\begin{figure}[!htbp]
    \centering
    \includegraphics[width=\columnwidth]{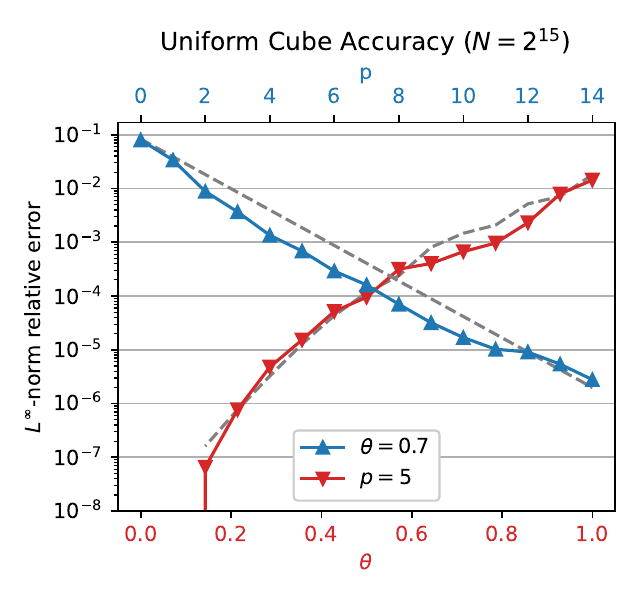}
    \caption{Relative error and theoretical scaling with effective theta $\theta_\text{eff}$ (gray dashed lines) for the uniform cube.}
    \label{fig:cube_errs}
\end{figure}

Next, we plot approximation errors for a fixed system size ($N=2^{15}$) and varying accuracy parameters $p$, $\theta$ (Figure \ref{fig:cube_errs}). Here, error again refers to the relative $L^\infty$-norm error compared to the analytic solution obtained via direct computation of the potential.
Both curves scale better than theoretically predicted by \eqref{eq:errbnd} since the effective theta is lower than the set bound. In other words, turning \eqref{eq:wellsep} into an equation and using it to find an effective $\theta$ for all interactions in the interaction lists returns e.g. $\theta_\text{eff}\approx 0.5 < 0.7 = \theta$. Such effective values are also used in the theoretical scaling of figure \ref{fig:cube_errs}, yielding the correct slopes.

Not plotted here are initialization/compilation time and the memory consumed for storing the hierarchy. The time required for initializing the algorithm is relatively high ($\approx 3.5$ minutes for $N=2^{20}$ and parameter set 1), which is primarily due to jit-compilation times in JAX. More specifically, the current implementation contains for-loops for the M2M/L2L transformations, all of which are unrolled in the compiled code which leads to long compilation times, particularly for high expansion orders $p$. On the other hand, the memory cost of the fully assembled hierarchy generally stays on the order of the memory required for the particle positions and charges. Here, we also remark that the interaction lists are the largest contribution to the memory usage and thus decreasing $\theta$ increases hierarchy memory usage proportionally to the increased number of M2L/direct interactions.

\begin{figure}[!htbp]
    \centering
    \includegraphics[width=0.75\columnwidth]{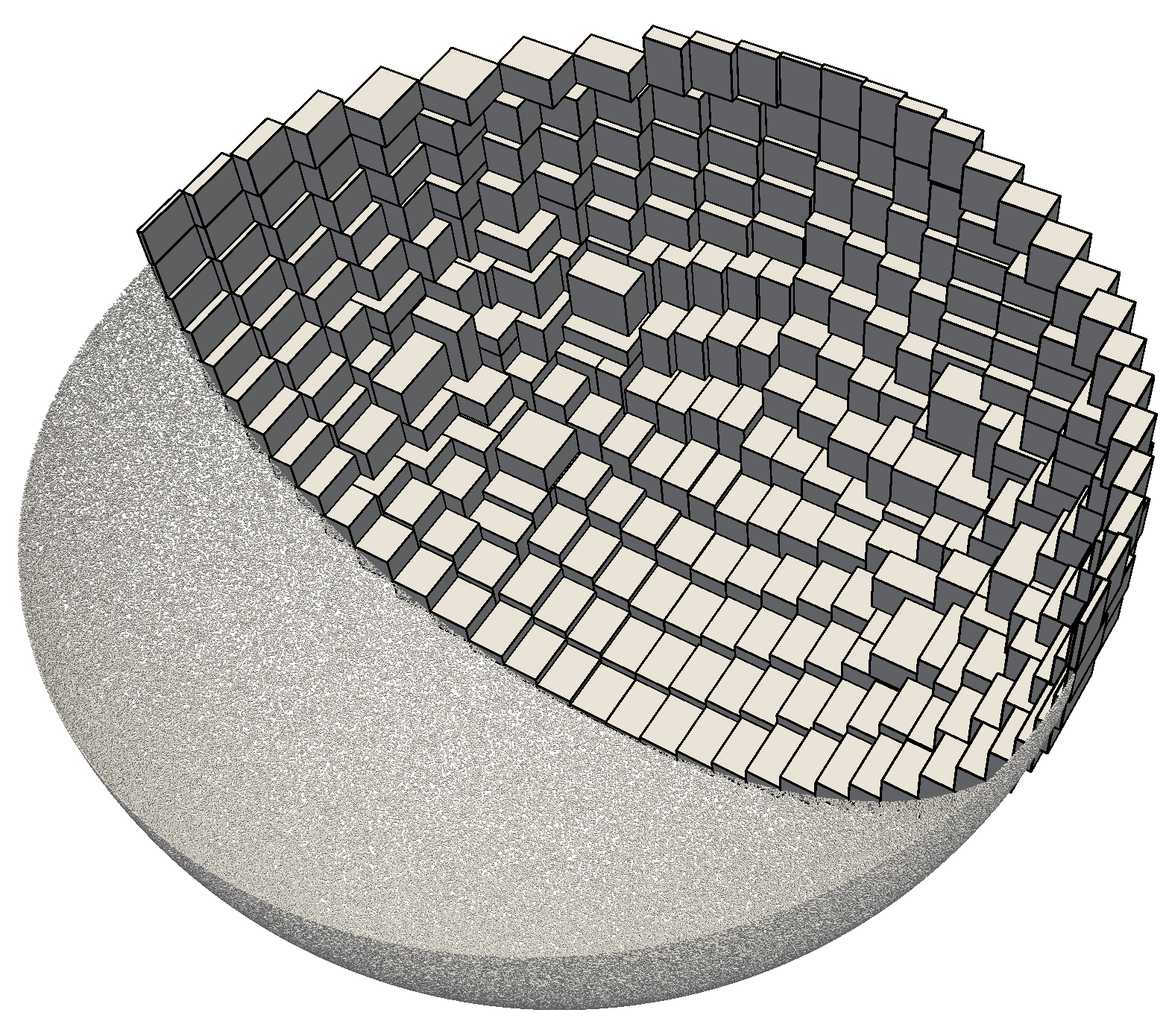}
    \caption{Level 5 of the sphere hierarchy (top right) and source points (bottom left) for $N=2^{20}$.}
    \label{fig:sp_hier}
\end{figure}

Next, we investigate points distributed uniformly on a sphere surface. This kind of distribution performs much better in adaptive FMM implementations. Figure \ref{fig:sp_hier} shows a cut of the generated hierarchy at level 5 and $N = 2^{20}$.

\begin{figure}[!htbp]
    \centering
    \includegraphics[width=\columnwidth]{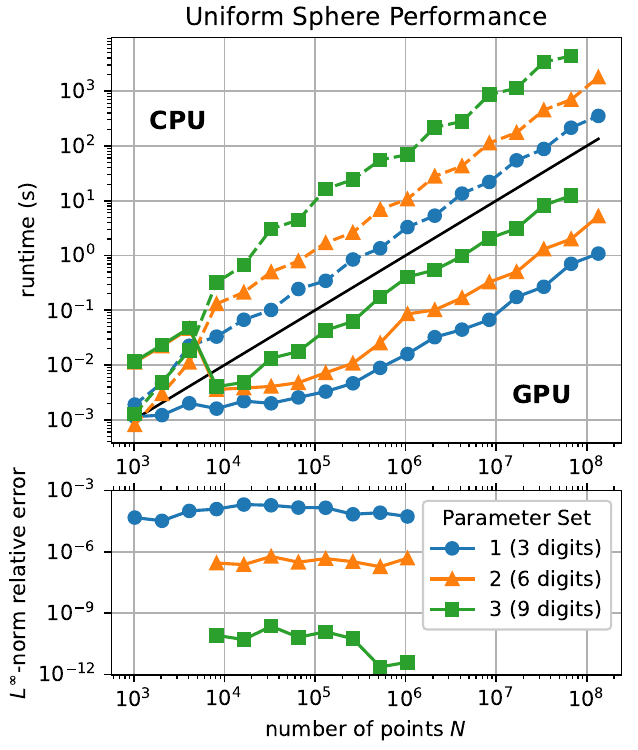}
    \caption{Uniform sphere performance on CPU (dashed lines) and GPU (solid lines).}
    \label{fig:sp_perf}
\end{figure}

\begin{figure}[!htbp]
    \centering
    \includegraphics[width=\columnwidth]{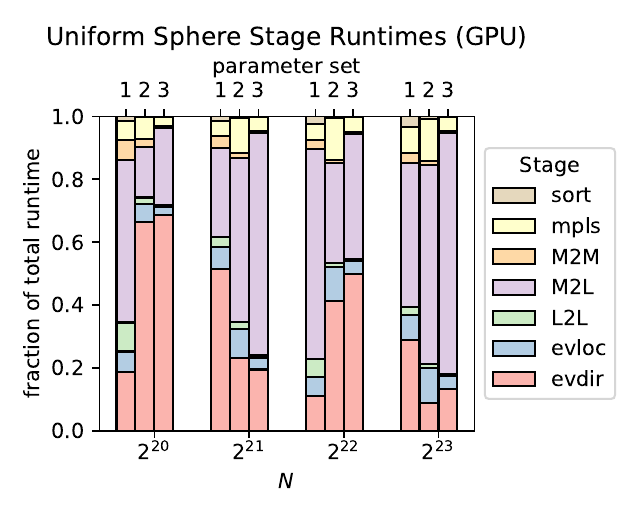}
    \caption{Uniform sphere stage timings at select problem sizes.}
    \label{fig:sp_substep}
\end{figure}

Similar to before, we have plotted runtimes and errors (Figure \ref{fig:sp_perf}) as well as stage timings (Figure \ref{fig:sp_substep}). Compared to the uniform cube distribution, runtimes even decrease slightly. This is due to the fact that boxes in the sphere hierarchy have comparatively fewer direct neighbors and as such can be handled at a lower level in the hierarchy, decreasing the total number of interactions that must be computed. In particular, this also decreases memory usage in the initialization, enabling potential evaluations of even larger systems. The decrease in interactions is also visible in the stage timings, where the non-M2L and non-evdir stages now take a larger fraction of the total runtime.

Finally, we compare the performance of jaxFMM to the highly optimized pvfmm \cite{malhotraPVFMMParallelKernel2015} implementation. We use the runtimes for the uniformly distributed cube at $N=2^{20}$ and additionally compute runtimes for the ellipsoid surface described in \cite{malhotraPVFMMParallelKernel2015} at $N=2^{20}$. Then, we compare to the corresponding results for $N=10^6$ in Tables 3 and 4 of the pvfmm publication (Table \ref{tab:compjaxfmm}). At roughly 3 digits of accuracy, jaxFMM on the GPU is about 4.5 times faster than pvfmm on 16 CPU threads for the uniform cube and about 10 times faster for the ellipsoid. We note that the 16 threads used for pvfmm equate to roughly 700 GFlops of peak single precision performance, compared to the 19.5 TFlops of an Nvidia A100 whose theoretical peak performance is thus about 28 times larger. At higher accuracies, jaxFMM is eventually outscaled by pvfmm's precomputed kernel-independent transformation operators with superior error scaling, except for the ellipsoid surface where jaxFMM is still faster at 9+ digits of accuracy. As previously discussed, we attribute the large difference in jaxFMM's CPU/GPU performance to poor CPU optimization within JAX. Overall, compared to pvfmm, jaxFMM's performance at high accuracies is limited. This is likely due to the analytic rotational transforms that exhibit worse error scaling and must be computed on-the-fly. On the other hand, considering the conciseness and simplicity of jaxFMM's code, it performs quite well at about 3 digits of accuracy, especially for the more non-uniform ellipsoid surface distribution.
\begin{table}[!htbp]
    \centering
    \begin{tabular}{|c|S[table-format=1.0e+1, round-mode = figures, round-precision = 1]S[table-format=2.1, scientific-notation = fixed, fixed-exponent = 0, round-mode = places, round-precision = 1]S[table-format=1.3, scientific-notation = fixed, fixed-exponent = 0, round-mode = places, round-precision = 3]|S[table-format=1.0e+1]S[table-format=1.2, scientific-notation = fixed, fixed-exponent = 0, round-mode = places, round-precision = 2]|}
    \cline{2-6}
    \multicolumn{1}{c|}{} & \multicolumn{3}{c|}{jaxFMM ($N=2^{20}$)} & \multicolumn{2}{c|}{pvfmm ($N=10^6$)} \\ \cline{2-6}
    \multicolumn{1}{c|}{}                    & {Error}    & {CPU ($n=1$)} & {GPU}     & {Error} & {CPU ($n=16$)} \\ \cline{1-6}
    \multirow{3}{*}{\shortstack{Uniform \\ Cube}}      & 8.1e-4  & 10.978416501078755   & 0.03446594695560634 & 5e-4  & 1.5e-1     \\
                                                       & 2.2e-6  & 38.78947082697414    & 0.25951942219398916 & 5e-6  & 2.5e-1     \\
                                                       & 1.3e-9  & 278.0675533900503    & 1.4244638569653034  & 7e-9  & 1.01e0      \\ \cline{1-6}
    \multirow{3}{*}{\shortstack{Ellipsoid \\ Surface}} & 6.8e-5  & 3.104563005035743    & 0.01358732208609581 & 1e-4  & 1.5e-1     \\
                                                       & 9.1e-7  & 6.445071031805128    & 0.06572941201739013 & 2e-6  & 2.4e-1     \\
                                                       & 1.3e-11 & 63.02898341114633    & 0.38247489114291966 & 8e-10 & 9.6e-1     \\ \cline{1-6}
    \end{tabular}
    \caption{Relative $L^\infty$-norm errors and runtimes (in seconds) of jaxFMM on CPU/GPU vs pvfmm \cite{malhotraPVFMMParallelKernel2015} on the CPU with $n=16$ threads.}
    \label{tab:compjaxfmm}
\end{table}

\subsection{Active Shielding}
\label{subsec:inverse_demo}
\begin{figure}[!htbp]
    \centering
    \includegraphics[width=\columnwidth]{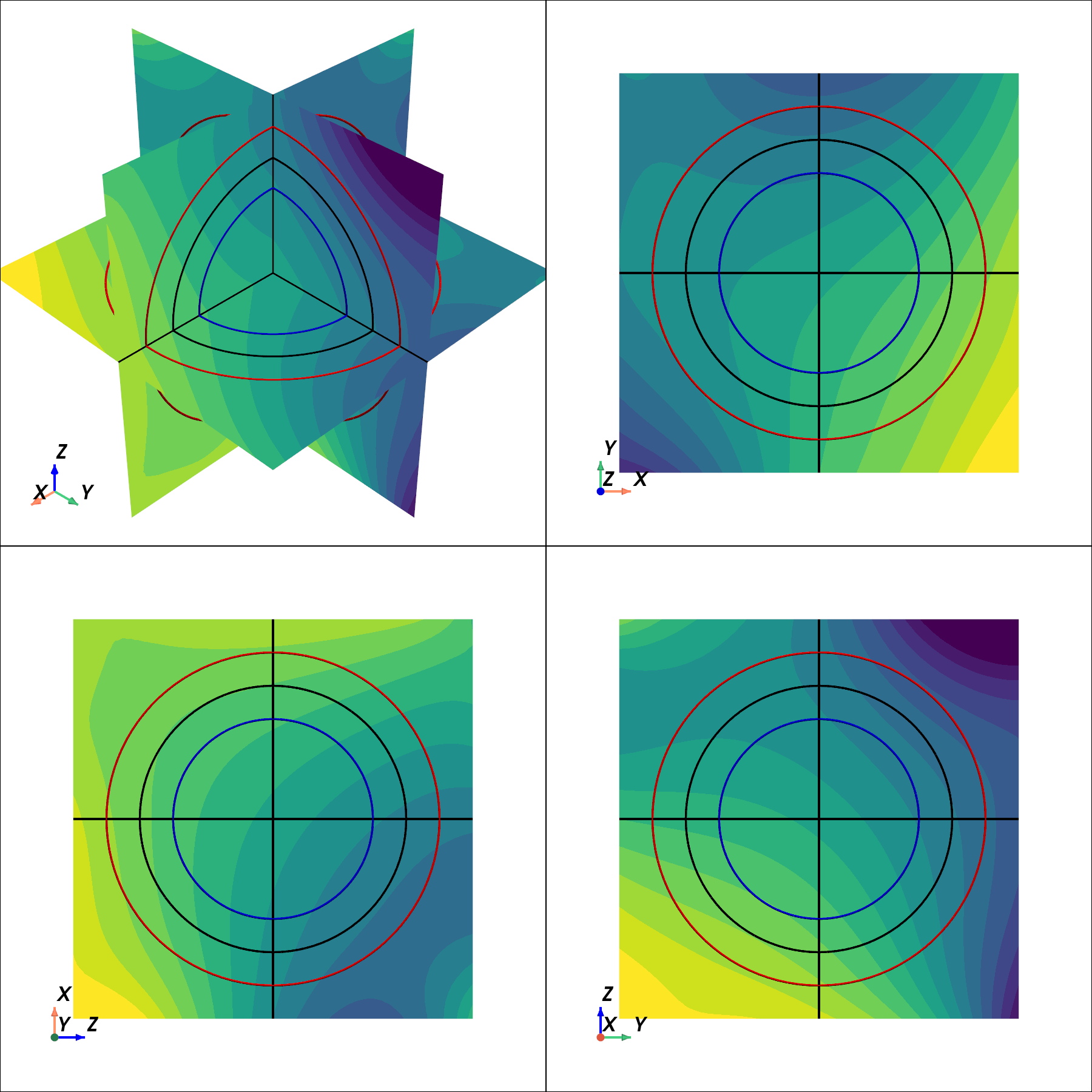}
    \caption{Potentials before optimizing shield charges (black circle) to minimize variance on the inside region (blue circle) and shield contribution on the outside (red circle).}
    \label{fig:inverse_before}
\end{figure}

\begin{figure}[!htbp]
    \centering
    \includegraphics[width=\columnwidth]{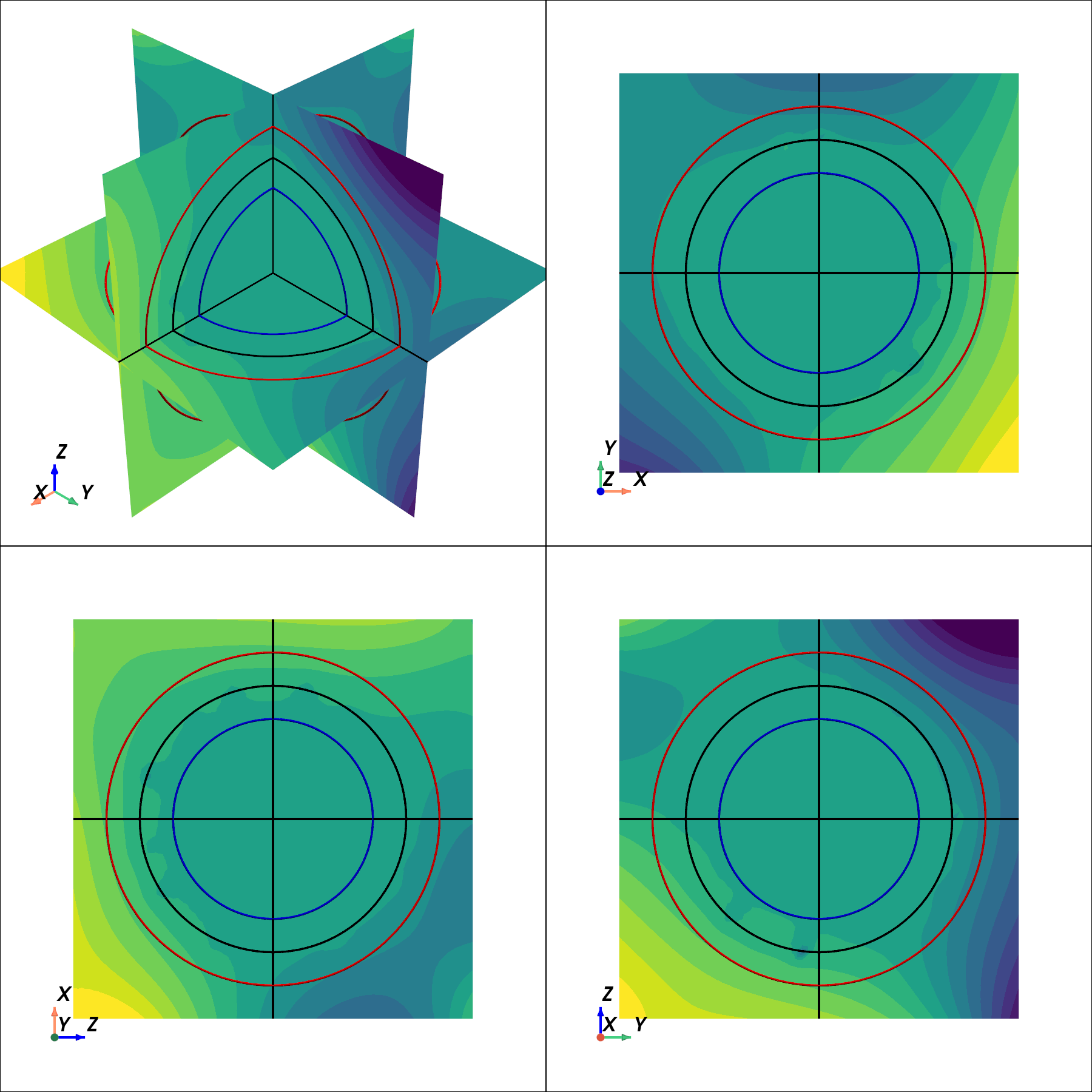}
    \caption{Potentials after optimizing shield charges (black circle) to minimize variance on the inside region (blue circle) and shield contribution on the outside (red circle).}
    \label{fig:inverse_after}
\end{figure}

To demonstrate the autodiff capabilities provided by JAX, we create a simple inverse design demo: $N_{\text{sh}}=128$ charges $q_i$ are spaced regularly on a sphere with fixed radius $R_{\text{sh}}=2.0$ to form an active shield against the potential introduced by $N_{\text{ext}}=256$ charges of uniformly random charge and position on a sphere of radius $R_{\text{ext}}=5.0$. The objective is to find shielding charges that minimize the variance of the potential at $N_{\text{in}}=1024$ points distributed evenly on a sphere of radius $R_{\text{in}} = 1.5 < R_{\text{sh}}$ in an effort to minimize the field. At the same time, the field on the outside should not be changed too drastically, which is ensured by checking the potential at $N_{\text{out}} = 1024$ evenly distributed points on a sphere of radius $R_{\text{out}} = 2.5 > R_{\text{sh}}$. Mathematically, we formulate the loss function as:
\begin{align}
    L_{\text{in}} &= \sum_{i=1}^{N_\text{in}} (\varphi^\text{(tot, in)}_i - \bar{\varphi}^\text{(tot, in)})^2 \\
    L_{\text{out}} &=  \sum_{i=1}^{N_\text{out}} (\varphi_i^\text{(sh, out)})^2 \\
    L_{\text{reg}} &= \sum_{i=1}^{N_\text{sh}} {q_i}^2 \\
    L_{\text{tot}} &= w_{\text{in}} L_{\text{in}} + w_{\text{out}} L_{\text{out}} + w_{\text{reg}} L_{\text{reg}}
\end{align}
where $\varphi_i^\text{(tot, in)}$ is the total potential at the sampling surface on the inside, $\bar{\varphi}^\text{(tot, in)}$ its mean, $\varphi_i^\text{(sh, out)}$ is the potential caused by the active shield on the outside sampling surface, $L_\text{reg}$ is an additional regularization term and $w_{\text{in}} = 1.0$, $w_{\text{out}} = 0.2$, $w_{\text{reg}} = 0.1$ are weights for the optimization. We stress here that while the optimization is trivial since the loss function only depends quadratically on the input charges, the strength of JAX lies within also being able to quickly construct and optimize much more complicated loss functions. For our demo, a simple gradient descent scheme suffices to quickly transform the potential on the inside from a highly nonuniform state (Fig. \ref{fig:inverse_before}) to a much more homogeneous state (Fig. \ref{fig:inverse_after}) with a comparatively small effect on the outside potential.

\section{Conclusions and Outlook}
\label{sec:conclusion}

In summary, we achieved an extremely concise, adaptive and highly parallel FMM implementation by leveraging JAX's just-in-time compilation. The code performs well at moderate accuracies, even for non-uniform charge distributions and performance can be fine-tuned through a variety of parameters. With JAX-intrinsic features like autodiff, exciting future applications such as coupling the FMM solver to complex inverse-design problems or machine-learning tasks can also be handled with ease, as demonstrated with a simple inverse-design problem.

While the current implementation requires only minimal padding to work within JAX, it suffers from long setup times due to jit-compilation, which are incurred every time the particle positions change. Additionally, the error scaling of the algorithm is inferior to state-of-the-art codes like pvfmm which, combined with the fact that the rotational transforms in jaxFMM must be computed on-the-fly for every single evaluation, prevents good performance at high accuracies. In an effort to avoid these limitations, a buffered uniform-refinement splitting algorithm with balance constraints is currently in development. This does introduce more padding, but has important advantages: The tree assembly can be fully jit-compiled, functions must only be recompiled for significant changes in the hierarchy and the regular grid structure allows for precomputed transformation operators which in turn enables performant kernel-independent and volume-FMM formulations. In particular, this will also allow computing gradients with respect to the charge geometry (particle positions) for shape optimization problems.

We remark that jaxFMM is primarily developed for fast stray-field evaluations in micromagnetic finite-element simulation codes, where the above drawbacks are not significant and jaxFMM already enables dynamic simulations of very large systems with tens of millions of elements. This is done by computing analytic corrections and point charges, the potential of which can be evaluated with Laplace-kernel FMM and converted into the magnetic field via a mass-lumping gradient procedure. A corresponding publication which will detail this process is forthcoming. JaxFMM is open-source on GitLab \cite{JaxFMMGitlabRepository2025}.

\begin{center}
    \textbf{Acknowledgements}\\
\end{center}

This research was funded in whole, or in part, by the Austrian Science Fund (FWF) P 34671 and I 6068. The computational results presented were achieved using the Vienna Scientific Cluster (VSC-5). For the purpose of open access, the author has applied a CC BY public copyright license to any Author Accepted Manuscript version arising from this submission.

\FloatBarrier
\clearpage

\bibliography{main}

\appendix

\section{Basis Functions}
\label{sec:basis_append}

\subsection{Regular Basis}
\label{sec:reg_rec}
\noindent For the regular basis defined as:
\begin{align*}
    R^m_n(\vec{r}) &= \alpha^m_n r^{n} Y_n^m (\theta,\varphi) \\
    \alpha^m_n &= \frac{(-1)^m}{(n+|m|)!} \\
    Y^m_n(\theta, \varphi) &= P^{|m|}_n(\cos(\theta)) \cdot e^{im\phi} \\
    P^m_n(x) &= (-1)^m (1-x^2)^{m/2} \frac{\diff^m}{\diff x^m}(P_n(x))
\end{align*}
where $Y^m_n$ are the (unnormalized) spherical harmonics, $P^m_n$ the associated Legendre functions and $P_n$ ordinary Legendre polynomials, we compute a real-valued equivalent $\hat{R}^m_n$ such that
\begin{equation*}
     \hat{R}^m_n = 
    \begin{cases} 
      \text{Re}(R^m_n) &  m\geq0 \\
      \text{Im}(R^{m}_n) & m<0
\end{cases}
\end{equation*}
with the following recursion relations:
\begin{equation*}
     \hat{R}_0^0 = 1, \quad \hat{R}_1^1 = \frac{x}{2}, \quad \hat{R}_1^{-1} = -\frac{y}{2}
\end{equation*}
For $n=2, \dots, p$:
\begin{align*}
    \hat{R}_n^n &= \frac{x \cdot \hat{R}_{n-1}^{n-1} + y \cdot \hat{R}^{1-n}_{n-1}}{2n} \\
    \hat{R}_n^{-n} &= \frac{x \cdot \hat{R}^{1-n}_{n-1} - y \cdot \hat{R}_{n-1}^{n-1}}{2n}
\end{align*}
For $n=0, \dots, p-1$:
\begin{equation*}
    \hat{R}^n_{n+1} = z \cdot \hat{R}^n_n, \quad \hat{R}^{-n}_{n+1} = z \cdot \hat{R}^{-n}_n
\end{equation*}
For $n=2, \dots, p$ and $m = 2-n, \dots, n-2$:
\begin{equation*}
    \hat{R}^m_n = \frac{(2n-1)\cdot z \cdot \hat{R}^m_{n-1} - r^2 \cdot \hat{R}^m_{n-2}}{(n-|m|)(n+|m|)}
\end{equation*}

\subsection{Singular Basis}
\label{sec:sing_rec}
\noindent Similarly to above, for the singular basis defined as:
\begin{align}
    S^m_n(\vec{r}) &= \beta^m_n r^{-(n+1)} Y_n^m (\theta,\varphi), \\
    \beta^m_n &= (-1)^m \cdot (n-|m|)!
\end{align}
we define a real-valued equivalent $\hat{S}^m_n$ such that
\begin{equation}
     \hat{S}^m_n = 
    \begin{cases} 
      \text{Re}(S^m_n) &  m\geq0 \\
      \text{Im}(S^{|m|}_n) & m<0
\end{cases}
\end{equation}
with the following recursion relations:
\begin{equation*}
     \hat{S}^0_0 = \frac{1}{r}, \quad \hat{S}^1_1 = \frac{x}{r^3}, \quad \hat{S}_1^{-1} = \frac{y}{r^3}
\end{equation*}
For $n=2, \dots, p$:
\begin{align*}
    \hat{S}^n_n &= (2n-1) \cdot \frac{x \cdot \hat{S}_{n-1}^{n-1} - y \cdot \hat{S}^{1-n}_{n-1}}{r^2} \\
    \hat{S}^{-n}_n &= (2n-1) \cdot \frac{x \cdot \hat{S}^{1-n}_{n-1} + y \cdot \hat{S}_{n-1}^{n-1}}{r^2}
\end{align*}
For $n=0, \dots, p-1$:
\begin{equation*}
    \hat{S}^{n}_{n+1} = (2n+1)\cdot \frac{z \cdot \hat{S}_n^n}{r^2}, \quad \hat{S}^{-n}_{n+1} = (2n+1)\cdot \frac{z \cdot \hat{S}_n^{-n}}{r^2}
\end{equation*}
For $n=2, \dots, p$ and $m = 2-n, \dots, n-2$:
\begin{equation*}
    \hat{S}^m_n = \frac{(2n-1)\cdot z \cdot \hat{S}^m_{n-1} - (n-1-m)(n-1+m)\cdot \hat{S}^m_{n-2}}{r^2}
\end{equation*}

\subsection{Regular Basis Gradient}
\label{sec:reg_grad_rec}
\noindent For field computations, it is useful to also know the gradient of the regular basis:
\begin{equation*}
     \nabla \hat{R}_0^0 = \vec{0}, \quad \nabla \hat{R}_1^1 = \frac{1}{2} \vec{e}_x, \quad \nabla \hat{R}_1^{-1} = - \frac{1}{2} \vec{e}_y
\end{equation*}
For $n=2, \dots, p$:
\begin{align*}
    \nabla \hat{R}_n^n &= -\frac{x \cdot \nabla \hat{R}_{n-1}^{n-1} + y \cdot \nabla \hat{R}^{1-n}_{n-1} + \hat{R}^{n-1}_{n-1} \cdot \vec{e}_x + \hat{R}^{1-n}_{n-1} \cdot \vec{e}_y}{2n} \\
    \nabla \hat{R}_n^{-n} &= \frac{x \cdot \nabla \hat{R}^{1-n}_{n-1} - y \cdot \nabla \hat{R}_{n-1}^{n-1} + \hat{R}^{1-n}_{n-1} \cdot \vec{e}_x - \hat{R}^{n-1}_{n-1} \cdot \vec{e}_y}{2n}
\end{align*}
For $n=0, \dots, p-1$:
\begin{equation*}
    \nabla \hat{R}^n_{n+1} = z \cdot \nabla \hat{R}^n_n + \hat{R}^n_n \cdot \vec{e}_z, \quad \nabla \hat{R}^{-n}_{n+1} = z \cdot \nabla \hat{R}^{-n}_n + \hat{R}^{-n}_n \cdot \vec{e}_z
\end{equation*}
For $n=2, \dots, p$ and $m = 2-n, \dots, n-2$:
\begin{align*}
    \nabla \hat{R}^m_n = [&(2n-1)(z \cdot \nabla \hat{R}^m_{n-1}+ \hat{R}^m_{n-1}\cdot \vec{e}_z)  \\
    &- r^2 \cdot \nabla \hat{R}^m_{n-2} - 2 \hat{R}^m_{n-2} \cdot \vec{r}] / (n^2-m^2)
\end{align*}

\section{Transformation Operators}
\label{sec:transf_ops}
In the following equations, we define the sign operator as:
\begin{equation*}
     \sgn(m) = 
    \begin{cases} 
      1 &  m\geq0 \\
      -1 & m<0
\end{cases}
\end{equation*}

\subsection{Azimuthal Rotations}
\noindent Multipole coefficients $C^m_n$ and local coefficients $D^m_n$ are rotated by an angle $\Delta\varphi$ around the $z$-axis as follows:
\begin{align*}
    C^m_n(\varphi + \Delta \varphi) = C^m_n(\varphi) \cos(m \Delta\varphi) - C^{-m}_n(\varphi) \sin(m \Delta\varphi) \\
    D^m_n(\varphi + \Delta \varphi) = D^m_n(\varphi) \cos(m \Delta\varphi) + D^{-m}_n(\varphi) \sin(m \Delta\varphi)
\end{align*}

\subsection{Polar Rotation}
\noindent Multipole expansions are rotated by a polar angle $\Delta\theta$ as follows:
\begin{align*}
    C^m_n(\theta + \Delta \theta) &= \sum_{k=0}^n  C^{\sgn(m)\cdot k}_n (\theta) \cdot N^m_{n,k} \cdot W^m_{n,k}(\Delta \theta)\\
    N^m_{n,k} &= \sqrt{\frac{(n-k)!(n+k)!}{(n-m)!(n+m)!}} \\
    W^m_{n,k}(\Delta \theta) &= 
    \begin{cases} 
        d^{|m|}_{n,k}(\Delta\theta) + \sgn(m)\cdot \\
        (-1)^k \cdot d^{|m|}_{n,-k} (\Delta\theta)  & |k| > 0\\
        d^{|m|}_{n,k}(\Delta\theta) & |k| = 0, |m| \geq 0 \\
        0   & |k|=0, |m| < 0
    \end{cases}
\end{align*}
where the Wigner-matrix coefficients $d^{m}_{n,k}$ are computed with fast Jacobi-recursions and symmetries found in \cite{wangEffectiveEfficientAlgorithm2022}. Similarly, for local expansions:
\begin{equation*}
    D^m_n(\theta + \Delta \theta) = \sum_{k=0}^n D^{\sgn(m)\cdot k}_n (\theta) \cdot \frac{W^m_{n,k}(\Delta \theta)}{N^m_{n,k}}
\end{equation*}

\subsection{Diagonal Transformations}
\noindent For transformations of length $r$ along the $z$-axis, $\vec{b} = (0,0,r)$, we shift multipole coefficients $C^m_n(\vec{a})$ and local coefficients $D^m_n(\vec{a})$ with diagonal transformations:
\begin{align*}
    C^m_n(\vec{a}+\vec{b}) &= \sum_{k=|m|}^n \frac{r^{(n-k)}}{(n-k)!} C^m_k(\vec{a}) &&\text{(M2M)}\\
    D^m_n(\vec{a}+\vec{b}) &= \sum_{k=|m|}^p \sgn(m)(-1)^{k+m} \frac{(n+k)!}{r^{(n+k+1)}} C^m_k(\vec{a}) &&\text{(M2L)}\\
    D^m_n(\vec{a}+\vec{b}) &= \sum_{k=n}^p \frac{(-r)^{(k-n)}}{(k-n)!} D^m_k(\vec{a}) &&\text{(L2L)}
\end{align*}
where the expansions are of order $p$, $n \leq p$ and $|m|\leq n$. We note that only the definition of the M2L operator differs from operators for complex-valued coefficients and only in the signs of the summation terms.
\end{document}